# Geant4 simulations of the lead fluoride calorimeter

A.A. Savchenko[a,*], A.A. Tishchenko[a], S.B. Dabagov[a,b], A. Anastasi[b,c], G. Venanzoni[b], M.N. Strikhanov[a] (et al.)

[a]National Research Nuclear University "MEPhI", Moscow, Russia
[b]Laboratori Nazionali di Frascati INFN, Frascati, Italy
[c]Dipartimento MIFT, Università di Messina, Italy

ABSTRACT

In this paper we simulate the emission by charged particles in complex structures with help of Geant4. We take into account Cherenkov radiation, transition radiation, bremsstrahlung, pair production and other accompanying processes. As an application we investigate the full size electromagnetic calorimeter for the muon g-2 experiment at Fermilab. A calorimeter module (24 are expected in the experiment) consists of a Delrin front panel for installation of the laser calibration system, 54 PbF$_2$ Cherenkov crystals wrapped by black paper, and silicon photo-multiplier sensors. We report here on a simulation of radiation from positrons striking the calorimeter system. We carry out the simulation using the Geant4 toolkit, which provides a complete set of tools for all areas of detector simulation: geometry, tracking, detector response, run, event and track management, and visualization. We consider Cherenkov photons expansion when a positron moves down through the calorimeter at the arbitrary angle of incidence. Both spectral and angular distributions of Cherenkov optical photons in different parts of the calorimeter system have been evaluated as well as the transition radiation and pre-shower distributions from the panel and from the Al vacuum chamber of the storage ring.

## 1. Introduction.

The BNL E821 experiment [1] has shown discrepancy between the Standard Model (SM) prediction of muon anomaly $\alpha_\mu$ and the measured one. The main goal of the new g-2 E989 experiment at Fermilab [2, 3] is to measure $\alpha_\mu$ with a fourfold improvement in accuracy with respect to the E821 experiment (0.54 part per million). The desired precision is 0.14 ppm. To achieve it, proper detectors are required, and simulation of processes occurring in the calorimeter modules plays a vital part. The aim of this paper is to investigate the electromagnetic calorimeter system using the Geant4 simulation toolkit, taking into account all the main processes accompanying the passage of a charged particle in a complex structure of a detector.

The paper is organized as follows. Sec. 2 describes the calorimeter system to be used in the E989 experiment. In Sec. 3 one can see preliminary results of our simulations divided into several steps, which differ in the setup geometry and in the primary particle energy. We end with summary and preliminary conclusions in Sec. 4.

## 2. The calorimeter system.

Twenty-four calorimeter modules placed on the inner radius of the muon storage ring provide precise measurements of the energy and hit time of positrons arising due to the muon decay ( $\mu^+ \rightarrow e^+ \bar{\nu}_\mu \nu_e$ ).

The highest-energy positrons appear when the muon spin and its momentum are co-directional. For the opposite directed muon spin and momentum we have positrons with lower energy. With retrieving this information, it is possible to evaluate the precession frequency of the muon spin. Together with a precise knowledge of the magnetic field, the muon anomaly can be obtained [2].

### 2.1. Experimental setup.

Each calorimeter module [4] consists of a Delrin front panel (with implantation of NBK-7 right-angle prisms) for installation of the laser calibration system [5, 6], 54 lead fluoride (PbF$_2$) crystals (6 high by 9 wide), wrapped by the black paper, and silicon photo-multiplier (SiPM) sensors [7]. The main PbF2 properties are its high density (7.77g/cm3), short radiation length (0.93cm), small Moliere radius (2.2cm), and transmittance range from 250 nm to 1100 nm. PbF2 is a Cherenkov radiator, and it does not produce as much light, as a scintillating crystal, but its light output allows one to have a necessary resolution for the positron energy reconstruction. Moreover, purely Cherenkov nature of the radiation makes high time resolution possible.

### 2.2. Simulation model.

In the simulation, the structure of calorimeter module is almost the same (see Fig. 1), but in addition to the above-mentioned structure we have a vacuum box, an aluminum (Al) plate in front of the Delrin panel, and an air gap between the front panel and the Al plate. At the experimental setup, the vacuum box and the Al plate are the vacuum chamber of the storage ring. The PbF2 Cherenkov radiators are supposed to have amorphous structure due to impossibility for Geant4 to work with the crystal structures directly. Although PbF2 is a crystal, it is possible to consider it as amorphous solid, taking into account its cubic crystal system and optical range of frequencies. We also have the difference in the sensitive part of the calorimeter module: air volumes, with an implementation of G4VSensitiveDetector class, represent SiPMs.

*Corresponding author.
E-mail address: aasavchenko1@mephi.ru

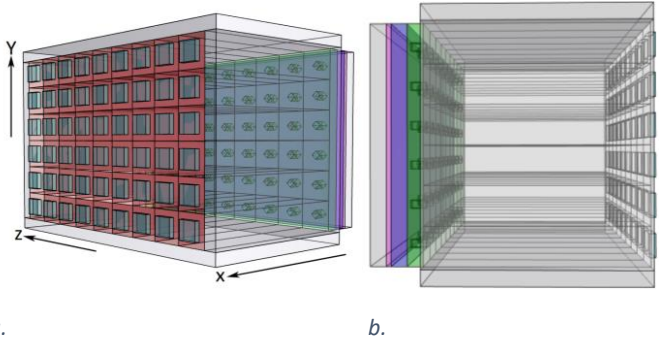

*a.*                *b.*

*Fig. 1 The Image of the calorimeter structure using Geant4 visualization tools. a. – 3D view of the calorimeter; b. (from left to right) the vacuum chamber, the Al plate, the air gap, the Delrin front panel (with implantation NBK-7 glass prisms), 54 PbF2 crystals, 54 sensitive air boxes.*

### 3. Geant4 simulations.

One can consider these simulations as an experiment carried out in the virtual laboratory. We perform our investigations using the Geant4 simulation toolkit, which has a width range of possibilities including simulation of the optical processes such as scintillations and the Cherenkov radiation. For use any of these processes, user has to switch it on in the special virtual class G4VUserPhysicsList and to create data tables of the material optical properties.

Now let us consider simulations for the different types of the geometry and the primary particle energy.

#### 3.1. Passage 2 GeV positrons through the vacuum chamber, Al plate and air gap.

We start from the simple geometry setting, specifying it in the virtual G4VUserDetectorConstruction class. For this step, we need the virtual model of the vacuum chamber, presented by the vacuum volume, 3 mm thick Al plate, and 1 cm air gap. Further, we simulate the passage of 1000 2 GeV positrons with the normal incidence through this structure. This allows us to investigate pre-shower distributions. In Fig. 2 one can see visualization of this step.

At the output, the simulation gives us information about angular and energy distributions of primary positrons (Fig. 3 and Fig. 4) and secondary particles (Figs. 5a,b) in the air gap. As is easy to see in Fig. 3, almost all primary positrons passed through Al plate have energy losses in the range of 50 – 100 MeV and, in very rare cases, have losses more than 100 MeV. Fig. 4 shows that primary positrons have a divergence less than 0.006 radian, which is important for the further simulation.

From output information about secondary particles, we find out that almost all "secondaries" are bremsstrahlung photons, which can further produce the Cherenkov radiation in $PbF_2$ crystals and in NBK-7 prisms. In addition, we see secondary electrons and transition radiation photons as well as photons with energy more than 200 MeV (see Fig. 5a), but they appear with a low probability.

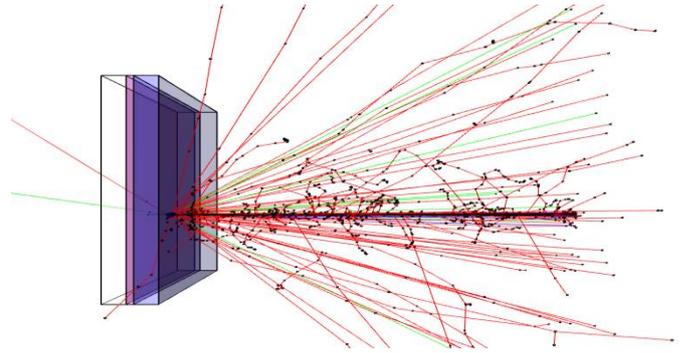

*Fig. 2 The shower of secondary particles (red – negative charged particles, blue – positive charged, green – neutral) after the passage of 1000 positrons through the Al plate.*

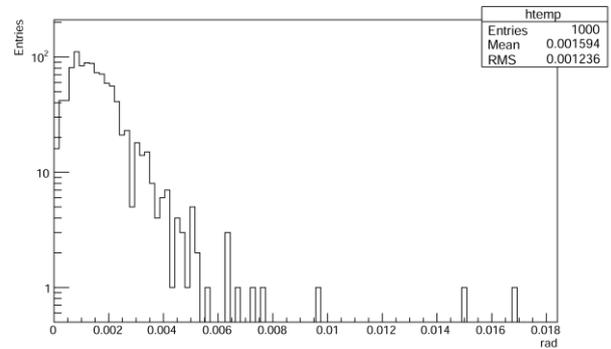

*Fig. 3 The angular distribution of primary positrons in the air gap after passing the Al plate. The angle is defined with respect to the primary momentum of positrons (1., 0., 0.).*

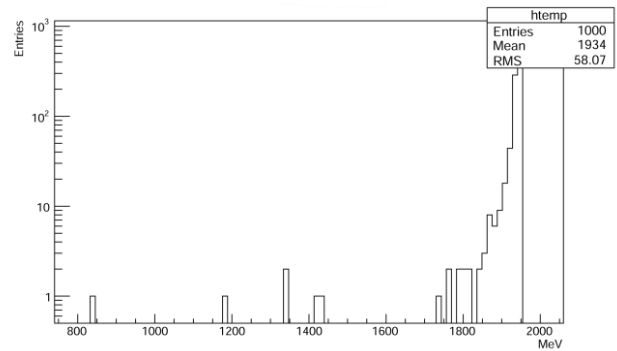

*Fig. 4 The energy distribution of primary positrons after losing the energy in the Al plate.*

Figs. 5a,b represent the energy distribution of secondary particles. One can see from Fig. 5b that the bulk part of "secondaries" has energy less than 20 MeV, which, however, is enough for further pair producing and for contributing to Cherenkov radiation. Moreover, the part of the "secondaries" has rather big angle with respect to the primary particles direction (see Fig. 2), which means further hitting by these particles not one but several $PbF_2$ crystals. In addition, there are photons with energies close to the primary positrons energy. Such particles can play significant role in the Cherenkov photon generation process.

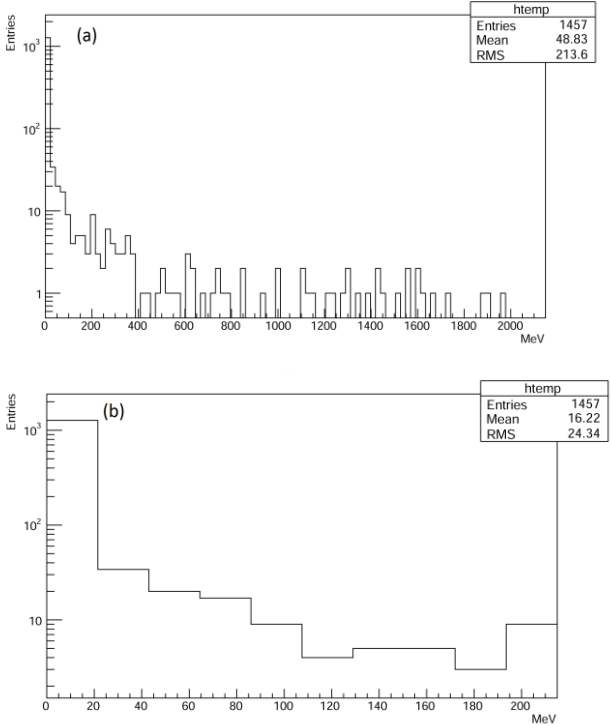

Fig. 5 The energy distribution of the secondary particle shower, which arise after positrons passing the Al plate.

### 3.2. Passage of the 2 GeV positrons through the vacuum chamber, Al plate, air gap, and Delrin front panel (with implantation of NBK-7 prisms).

Let us consider the simulation of the 1000 positrons passage through the above-simulated structure and the 1 cm thick Delrin front panel (with an implantation of NBK-7 right-angle prisms). Fig. 6 visualizes the spatial distribution of primary positrons (blue) and the secondary particles shower.

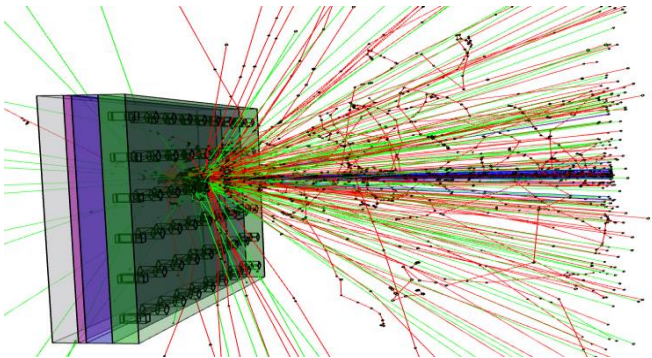

Fig. 6 The shower of secondary particles (red – negative charged particles, blue – positive charged, green – neutral) after passage of 1000 positrons through the Al plate and the Delrin front panel.

The main result of this step is the investigation of Cherenkov photons arising in the NBK-7 prisms. As one can see from Fig. 7, Cherenkov radiation occurs in NBK-7 prisms in the photon energy range from 1.2 eV (1033 nm) to 3.5 eV (354 nm). The amount of such photons is restricted by the small size of the NBK-7 prism (5x5x5 mm³), due to the dependence of Cherenkov radiation on radiator size (length, to be precise). Moreover, the total internal reflection angle for NBK-7 $\left(\theta_c \approx 41°\right)$ is less than Cherenkov one $\left(\theta_{cher} \approx 49°\right)$ for 2 GeV positrons passing through NBK-7. Therefore, for normal incidence positrons, Cherenkov photons cannot go out of the prism and do not contribute to the total amount of Cherenkov photons registered by SiPMs in the next steps of the simulation.

As in the previous step, we evaluate energy distributions of the secondary particles (Fig. 8) and primary positrons (Fig. 9). One can notice that the energy distributions both for primary and secondary particles become wider due to the positron passage through rather thick plastic panel. In addition, we see that secondary particles (Fig. 6) have the bigger divergence comparing with the simulation without the Delrin front panel (Fig. 2).

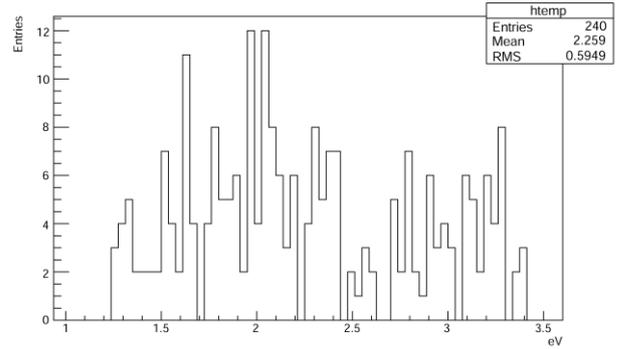

Fig. 7 The energy distribution of Cherenkov photons arising in the NBK-7 right-angle prisms.

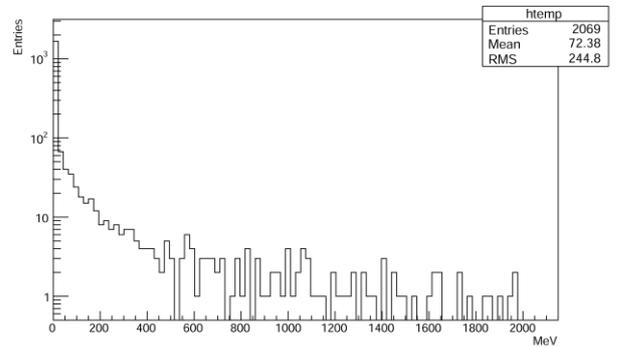

Fig. 8 The energy distribution of the secondary particle shower, which arise after positrons passage through the Delrin front panel.

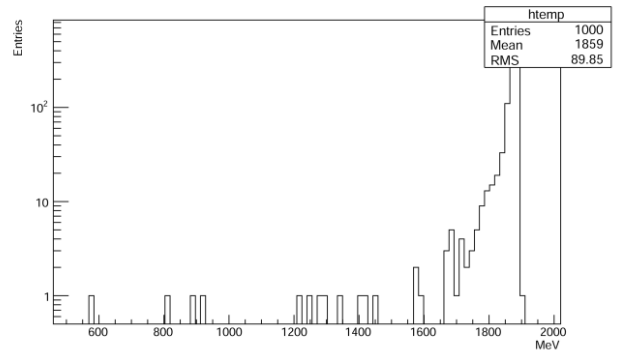

Fig. 9 The energy distribution of primary positrons after losing the energy in the Al plate and Delrin front panel.

We also consider an additional effect: transition radiation from the Al plate and the Delrin front panel. This effect is very important, because the transition radiation is a precise tool for particle energy measurement due to its strong dependence on the Lorentz factor.

For the g-2 experiment, it could be an alternative way for positron detecting.

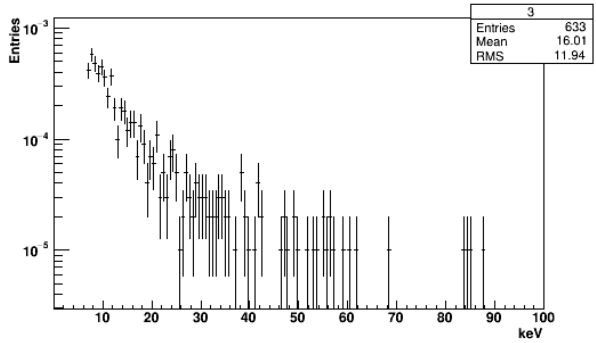

Fig. 10 The energy spectrum of transition radiation photons arising in the Al plate after passage of 100000 positrons.

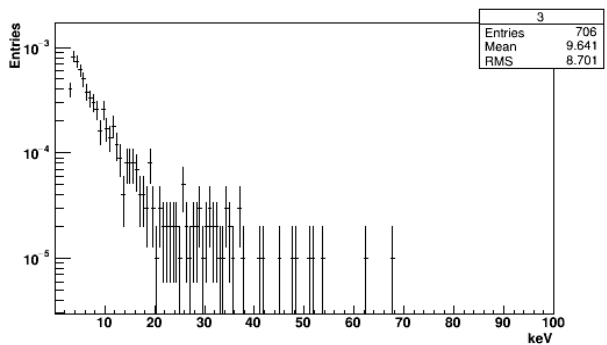

Fig. 11 The energy spectrum of transition radiation photons arising in the Delrin front panel after passage of 100000 positrons.

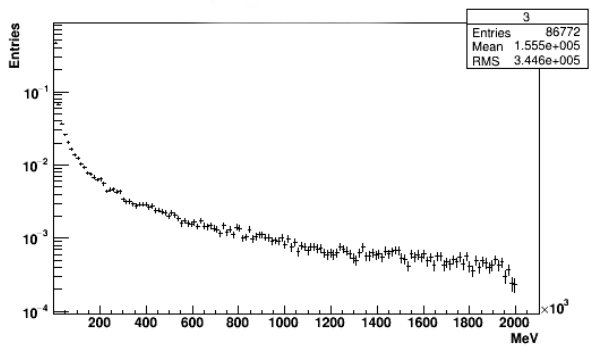

Fig. 12 The energy distribution of the transition radiation and the bremsstrahlung measured after passage of $10^5$ positrons through the Al plate and the Delrin front panel.

Figs. 10, 11 show the energy distribution of transition radiation. As on can see, there is considerable amount of TR photons, which is possible to use for alternative positron detecting system. Both the bremsstrahlung and transition radiation are measured (Fig. 12) for the case of the $10^5$ positrons passing the aluminum and Delrin panel.

### 3.3. The crystals wrapping.

In the experimental setup, PbF$_2$ crystals are wrapped by Millipore black paper. It is needed for separation of Cherenkov light between neighbored crystals. The Millipore paper is a polyvinylidene fluoride membrane with 0.45 um pores. Due to its structure, the Millipore paper works like a diffusive mirror.

For Geant4 optical processes simulation, properly chosen wrapping is very important. If we do not use the wrapping, Geant4 will consider 54 crystals like one volume. It is possible to use several types of wrapping. First, the G4OpticalSurface class (a logical surface or a logical skin) can be used. The logical skin works perfect, but it covers all surfaces, which does not allow input and output of optical photons. With using the logical surface, one can cover surfaces, which should be covered. But it does not work between volumes with the same refractive index, such as our crystals. Second, one can create the volumes with special optical features and place them between crystals. This kind of wrapping does not depend on the crystal refractive index, but in order to use properly separating volumes one have to create material optical properties tables. Without these tables, a chosen volume material can only absorb photons but not refract and reflect them. The possibility of light separation between identical volumes makes us choose the latter method. Nevertheless, it also has disadvantage such as increasing of simulation time because of bigger amount of volumes and creation by Geant4 additional particle tracks inside separation volumes.

### 3.4. Simulation of the full-size calorimeter module.

Here, we complete the structure mentioned in Sec 3.2. with 54 crystals and 54 sensitive vacuum boxes. Since this moment, one can consider it as a full-size electromagnetic calorimeter. Firstly, for geometry checking, we decide to carry out preliminary simulations for positrons with energies less than 2 GeV.

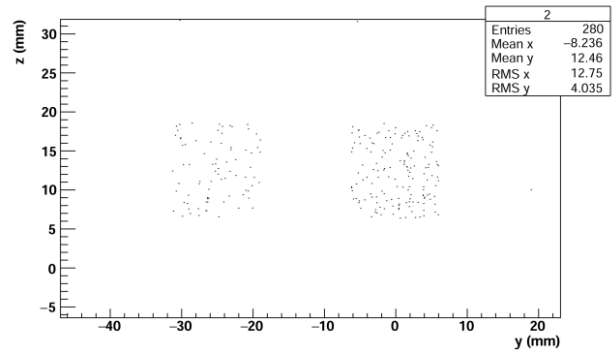

Fig. 13 Registered by sensitive boxes Cherenkov light signal from the 50 MeV positron.

We test the calorimeter system with 50 MeV positrons and find out that positrons with such energy are deflected in the Delrin front panel and cross several neighbored crystals. In Fig. 13 one can observe several signals from only one primary positron. In this case, it is difficult to the reconstruct positron energy, since for the detector these signals are seemed to be from two different particles.

Sensitive volumes allow us not only to see photon spatial distribution but also to measure the Cherenkov radiation energy distribution (Fig. 14). The transmission feature of the PbF2 crystals restricts the range of the Cherenkov photons energy, which, however, does not matter for the experiment, since SiPM photo-detection-efficiency is in the range from 340 nm (3.6 eV) to 900 nm (1.37 eV) and it is peaked around 420 nm (3 eV).

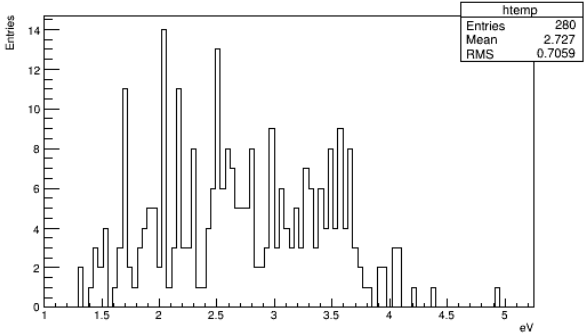

Fig. 14 The energy distribution of the registered Cherenkov photons by sensitive detectors for the primary 50 MeV positron.

After testing the calorimeter working ability with low energy positrons, we proceed to the main investigations, which are dedicated to the simulation of the 2 GeV positron passage through the full-size calorimeter module. We carry out simulations using NRNU MEPhI high-performance computing center.

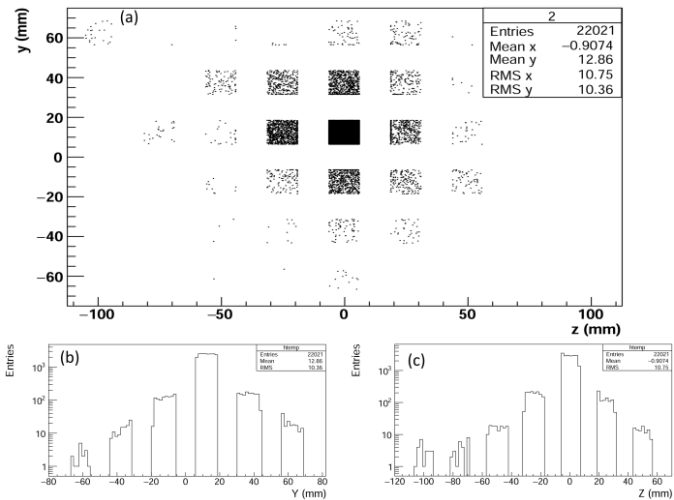

Fig. 15 The spatial distribution of Cherenkov photons registered by sensitive volumes.

As it is for 50 MeV case, sensitive detectors give us information about the spatial distribution of Cherenkov photons (Figs. 15a,b,c), about the photon hitting time, which talks us about the time resolution of the calorimeter, the Cherenkov photons energy (Figs. 17, 18) and angular (Fig. 19) distributions.

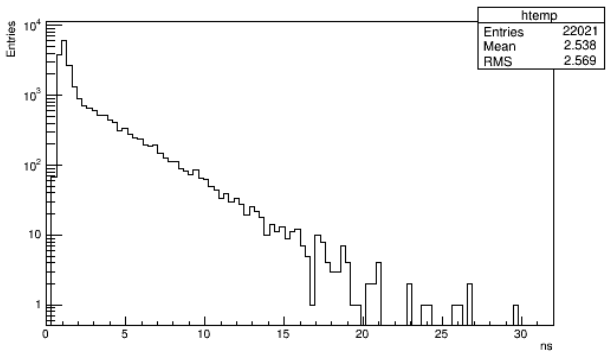

Fig. 16 The time of the hit registration by sensitive volumes. It starts from the generation of the primary vertex.

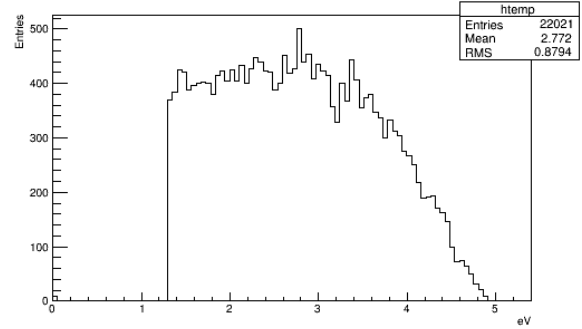

Fig. 17 The energy distribution of the Cherenkov photons registered by sensitive detectors.

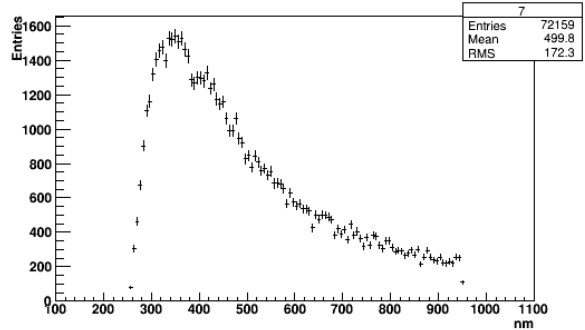

Fig. 18 The wavelength distribution of Cherenkov photons at the back surface of the $PbF_2$ crystal.

Figs. 15b,c show us the spatial distributions of optical photons. This information is very important for the experiment. Indeed, the more photons are uniformly distributed on back surface of the crystal, the more complete will be information after the digitization of the SiPM signal, but in Fig 15a one can see clearly that the active SiPM area is one fourth of the rear crystal surface. Moreover, in spatial distribution one can see the main signal from the crystal hit by the positron and penetration of Cherenkov photons to several crystals. Investigations of this expansion show that it can be caused by scattering of the high-energy secondary particles (see Fig. 19). The crystal wrapping does not allow Cherenkov photons crossing borders between crystals; for high-energy photons, however, our wrapping is transparent, which allow them to cross borders and generate Cherenkov radiation in the neighbored crystals. For the detector this effect could mean incompleteness of the output information due to the energy spreading, but the bulk part of photons is concentrated in crystal hit by the positron, which means that one can consider other photons as a background.

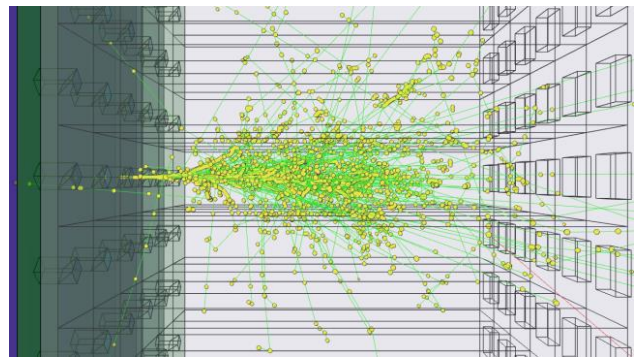

Fig. 19 The shower of secondary particles in $PbF_2$ crystal hit by 2 GeV positron. Green tracks are trajectories of neutral particles, such as X-ray and gamma photons (visualization of Cherenkov photons is turned off).

As to the completeness of the information, the energy distributions (see Figs. 17, 18) show that Cherenkov photons have energies in the range from 240 nm to 1000. Along with it, SiPM sensitivity is between 340 nm and 900 nm, which means that SiPMs do not register sixth part of Cherenkov photons, meaning the energy. This fact can introduce an uncertainty in the final positron energy reconstruction. As one can see in Fig. 16, the most part of the registered photons hit the sensitive volume 3 ns after the primary vertex generation, that implies a nanosecond scale resolution of the calorimeter. Fig. 20 presents the angular distribution, which is in a good agreement with that preliminary evaluated for the PbF2 crystal.

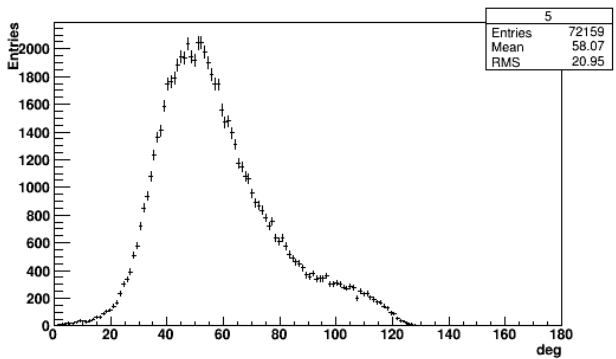

Fig. 20 The angular distribution of Cherenkov photons at the back surface of the PbF2 crystal (the Cherenkov angle for the PbF2 = 56.83 degrees). The angle is defined with respect to the primary momentum of positrons (1., 0., 0.).

4. Summary

In this paper we investigated the full-size electromagnetic calorimeter for the muon g-2 experiment. It is the current work, which means, that we can make only preliminary, but rather important conclusions. We considered pre-shower distributions both for the aluminum exit of the muon storage ring and the Delrin front panel and shown that due to its high energy distributions these pre-showers contribute to the radiation yield. The main mechanism for generation of the pre-showers is proved to be bremsstrahlung.

During the Delrin front panel simulation, we showed that Cherenkov radiation can arise in the NBK-7 prisms, but due to prism features, photons cannot go out, and consequently, these photons do not contribute in the total amount of photons registered by SiPMs.

Simulating the full-size calorimeter module we faced the fact that the primary positron with the energy less than 50 MeV could be deflected in the Delrin front panel and hit not only one crystal but several. In the latter case, we have several signals with the same intensity from one positron, which means impossibility of the positron energy reconstruction.

At last, we simulated 2GeV positron passage through the full-size calorimeter. Evaluating distributions of Cherenkov photons, we show that due to the SiPM sensitivity range at the experiment one cannot register sixth part of Cherenkov photons. Due to this fact, one can have the uncertainty in the positron energy reconstruction. Along with that, we demonstrated the excellent spatial distribution of photons at the crystal back surface, which is certain important for the precise measurement of the primary positron energy. We evaluated both energy and angular distributions as well as the time distribution, which shows nanoscale time resolution.

Also, the results of this paper provide the basis for our Geant4 simulations of the transition radiation from regular radiators in the framework of the ATLAS project.

Acknowledgements

This work was supported under grant RSF 16-12-10277. The opportunity to perform computer simulations at NRNU MEPhI high-performance computing center is also gratefully acknowledged.